\documentclass[11pt,a4paper]{article}
\pdfoutput=1
\usepackage{jheppub}
\usepackage{amssymb}
\usepackage{graphics}
\usepackage{bm}
\usepackage{graphicx,bm}
\usepackage{amsmath,amssymb,mathrsfs,verbatim,subfigure}

\renewcommand{\thanks}[1]{\footnote{#1}}

\newcommand{\bea}{\begin{eqnarray}}
\newcommand{\eea}{\end{eqnarray}}
\newcommand{\ee}{\end{equation}}
\newcommand{\be}{\begin{equation}}

\def\det{{\rm det}}

\def\p{\partial}

\def\l({\left(}
\def\r){\right)}

\title{%Inverse magnetic catalysis in hard-wall AdS/QCD and flavored $\mathcal{N}=4$ SYM on $R^3\times S^1$
Inverse magnetic catalysis in holographic models of QCD}

\author{Kiminad A. Mamo}

 \affiliation{Department of Physics, University of Illinois, Chicago, Illinois 60607, USA}

\emailAdd{kabebe2@uic.edu}

\abstract{
We study the effect of magnetic field $B$ on the critical temperature $T_{c}$ of the confinement-deconfinement phase transition in hard-wall AdS/QCD, and holographic duals of flavored and unflavored $\mathcal{N}=4$ super-Yang Mills theories on $\mathbb{R}^3\times \rm S^1$. For all of the holographic models, we find that $T_{c}(B)$ decreases with increasing magnetic field $B\ll T^2$, consistent with the \textit{inverse magnetic catalysis} recently observed in lattice QCD for $B\lesssim 1~GeV^2$. We also predict that, for large magnetic field $B\gg T^2$, the critical temperature $T_{c}(B)$, eventually, starts to increase with increasing magnetic field $B\gg T^2$ and asymptotes to a constant value.}

\arxivnumber{}
\begin{document}

\maketitle

\section{Introduction}
Recently, the study of the QCD phase diagram for magnetic field $B$ has attracted considerable attention \cite{Gusynin:1994re,Miransky:2002rp,Mizher:2010zb,Fraga:2008um,Gatto:2010pt,Gatto:2010qs,Osipov:2007je,Kashiwa:2011js,Klimenko:1992ch,Alexandre:2000yf,Filev:2007gb,Albash:2007bk,Alam:2012fw,Johnson:2008vna,Bergman:2008sg,Evans:2010xs,Preis:2010cq,Ballon-Bayona:2013cta,Bali:2011qj}, see \cite{Kharzeev:2012ph} for a review. The main motivation for studying the QCD phase diagram under external magnetic field $B$ stems from the fact that strong magnetic field $B$ is produced in heavy ion collisions experiments at RHIC $eB\sim 0.01~GeV^2$ and LHC $eB\sim 0.25~GeV^2$ \cite{Skokov:2009qp}, due to the charged spectator particles, which has interesting effects on the quark-gluon plasma created during these heavy ion collision experiments \cite{Kharzeev:2007jp,Fukushima:2008xe,Kharzeev:2004ey,Voloshin:2008jx,Abelev:2009uh,Selyuzhenkov:2011xq}, see \cite{Kharzeev:2012ph} for a review. A strong magnetic field $eB\sim 4~GeV^2$ is also produced during the electroweak phase transition of the early Universe \cite{Vachaspati:1991nm}, and relatively weaker magnetic field $eB\sim 1~MeV^2$ is produced in the interior of dense neutron stars \cite{Duncan:1992hi}.

Another motivation comes from the fact that the study of the QCD phase diagram with magnetic field $B$ is amenable to numerical simulations of QCD on the lattice, without facing the sign problem of lattice QCD that exist in the case of non-zero baryon chemical potential $\mu_{B}$, creating an opportunity to compare the holographic and low energy effective models of QCD directly with QCD itself.

Regarding the study of the QCD phase diagram for magnetic field $B$, most of the models for QCD \cite{Gusynin:1994re,Miransky:2002rp,Mizher:2010zb,Fraga:2008um,Gatto:2010pt,Gatto:2010qs,Osipov:2007je,Kashiwa:2011js,Klimenko:1992ch,Alexandre:2000yf}, including the holographic ones \cite{Filev:2007gb,Albash:2007bk,Alam:2012fw,Johnson:2008vna,Bergman:2008sg,Evans:2010xs}, have studied chiral-symmetry-restoration transition and have predicted that the critical temperature $T_{c}$ of the transition increases with increasing magnetic field $B$ at zero chemical potential $\mu=0$. This enhancing effect of the magnetic field $B$ on the critical temperature $T_{c}$ has been termed \textit{magnetic catalysis}. However, recent lattice QCD result \cite{Bali:2011qj} has indicated the opposite effect, that is, the critical temperature $T_{c}$ decreases with increasing magnetic field $B$, for $B\lesssim 1~GeV^2$ and zero chemical potential $\mu=0$. This inhibiting effect of the magnetic field $B$ on the critical temperature $T_{c}$ has been termed \textit{inverse magnetic catalysis}.

Even though, the recent lattice QCD result \cite{Bali:2011qj} has also indicated that the confinement-deconfinement and chiral symmetry breaking phase transitions occur at the same critical temperature $T_{c}(B)$ at least for $B\lesssim 1~GeV^2$, most holographic calculations so far \cite{Filev:2007gb,Albash:2007bk,Alam:2012fw,Johnson:2008vna,Bergman:2008sg,Evans:2010xs,Preis:2010cq} have been concerned only with $T_{c}(B)$ of the chiral symmetry breaking phase transition.% and the $T_{c}(B)$ of confinement-deconfinement phase transition has not been studied. %In addition, none of them has %given direct holographic calculation of $T_{c}(B)$ which has shown \textit{inverse magnetic catalysis} at zero chemical potential $\mu=0$.

However, recently, reference \cite{Ballon-Bayona:2013cta}, inspired by the recent lattice QCD result \cite{Bali:2011qj}, has a priori assumed confinement and chiral symmetry breaking transitions to occur at the same critical temperature $T_{c}$ in Sakai-Sugimoto model, and has argued that, in this case, $T_{c}(B)$ must be a decreasing function of $B$, consistent with the recent lattice QCD result \cite{Bali:2011qj}, but has not provided a direct computation of $T_{c}(B)$.

In this paper, we give a direct computation of the critical temperature $T_{c}(B)$ of the confinement-deconfinement phase transition in hard-wall AdS/QCD, and holographic duals of flavored and unflavored $\mathcal{N}=4$ SYM on $\mathbb{R}^3\times \rm S^1$ where $S^1$ is a circle of length $l$ in one of the spatial directions. (Note that, at finite temperature $T$, $\mathbb{R}^3$ is really $S_{\tau}^{1}\times \mathbb{R}^2$ where $S_{\tau}^{1}$ is the thermal circle with length $\frac{1}{T}$.) Also, note that, since the fermions of both the flavored and unflavored $\mathcal{N}=4$ SYM on $\mathbb{R}^3\times \rm S^1$ obey antiperiodic boundary conditions around the circle $\rm S^1$, they acquire a tree-level mass $m\sim\frac{1}{l}$. The scalars are periodic around the circle, hence they acquire masses only at the quantum
level through their couplings to the fermions \cite{CasalderreySolana:2011us}. The gluons, however, do not acquire masses, therefore, at low-energy, both flavored and unflavored $\mathcal{N}=4$ SYM on $\mathbb{R}^3\times \rm S^1$ reduce to pure 3D Yang-Mills theory.

It is well known that both flavored and unflavored $\mathcal{N}=4$ super-Yang Mills theories (SYM) on flat spacetime $\mathbb{R}^4$ are not confining gauge theories. However, they can be made confining in the large-$N_{c}$ limit by placing them on a compact space with length $l$, and the confinement-deconfinement phase transition occurs at critical temperature $T_{c}=\frac{1}{l}$ \cite{Witten:1998zw,Surya:2001vj,Sonnenschein:2000qm}, see \cite{CasalderreySolana:2011us,Natsuume:2014sfa} for a review. In our case, the compact space is $\mathbb{R}^3\times \rm S^1$, that is, we compactify one of the spatial dimensions into a circle of length $l$.

The confinement-deconfinement phase transition both in flavored and unflavored $\mathcal{N}=4$ SYM on $\mathbb{R}^3\times \rm S^1$ is holographically modeled by a phase transition between a black hole solution with radius of horizon $r=r_{h}$, and $AdS_{5}$-soliton solution which smoothly ends at $r=r_0$. However, to study the confinement-deconfinement phase transition in QCD on $\mathbb{R}^4$ at strong coupling, we use the hard-wall AdS/QCD model where the confinement-deconfinement phase transition, of QCD on $\mathbb{R}^4$, is holographically modeled by a phase transition between a black hole solution with radius of horizon $r=r_{h}$, and thermal-$AdS_{5}$ solution with hard-wall IR cut-off $r=r_{0}$.

We derive the corresponding thermal-$AdS_{5}$ solution which is the holographic dual to the confined phase of QCD on $\mathbb{R}^4$ by starting from a black hole solution, which corresponds to the deconfined phase of strongly coupled QCD on $\mathbb{R}^4$, by setting the mass of the black hole to zero \cite{Lee:2009bya}. And, we derive the corresponding $AdS_{5}$-soliton solution, which is the holographic dual to the confined phase of flavored and unflavored $\mathcal{N}=4$ SYM on $\mathbb{R}^3\times \rm S^1$, by "double Wick rotating" a black hole solution \cite{CasalderreySolana:2011us,Natsuume:2014sfa}.

In this paper, we use two black hole solutions in the presence of constant magnetic field $B$. First, we use the black hole solution in the presence of constant magnetic field $B\ll T^2$ found in \cite{D'Hoker:2009bc} to study the confinement-deconfinement phase transition in strongly coupled QCD on $\mathbb{R}^4$ and unflavored $\mathcal{N}=4$ SYM on $\mathbb{R}^3\times \rm S^1$. Then, we use the black hole solution in the presence of constant magnetic field $B$, including the backreaction of $N_{f}$ flavor or D7-branes for $N_{f}\ll N_{c}$, found in \cite{Ammon:2012qs} to study the confinement-deconfinement phase transition in flavored $\mathcal{N}=4$ SYM on $\mathbb{R}^3\times \rm S^1$.

The effect of magnetic field $B$ on different observables has also been studied in \cite{Basar:2012gh,Mamo:2013efa,Arciniega:2013dqa,Critelli:2014kra,Rougemont:2014efa} using the backreacted black hole solution of \cite{D'Hoker:2009bc} without flavor D7-branes.

Depending on the specific holographic models to QCD, various length and energy scales appear throughout this paper. Some of the relevant length and energy scales are: the radius of the $AdS_{5}$ spacetime $L$ which we set to $L=1$, the radius of the black hole horizon $r_{h}$ which is related to the Hawking temperature $T_{H}$ of the black hole (which is dual to the field theory temperature $T=T_{H}$), the radial position of the canonical singularity of the $AdS_{5}$-soliton $r_{0}=\pi T_{c}(B=0)=\pi\times 0.175~GeV=0.55~GeV$ for flavored and unflavored $\mathcal{N}=4$ SYM on $\mathbb{R}^3\times \rm S^1$, the radial position of the hard-wall $r_{0}=\frac{m_{\rho}}{2.405}=0.323~GeV$ in the thermal-$AdS_{5}$ solution for the hard-wall AdS/QCD, and an external magnetic field $B$ in the range of $0-0.35~GeV^2$ for the hard-wall AdS/QCD model and $0-4.2~GeV^2$ for the flavored $\mathcal{N}=4$ SYM on $\mathbb{R}^3\times \rm S^1$.

%Equipped with these backreacted black hole, thermal-$AdS_{5}$, and $AdS_{5}$-s oliton geometries, we give a direct computation of $T_{c}(B)$ in hard-wall %AdS/QCD \cite{Erlich:2005qh,deTeramond:2005su}, and holographic duals of flavored and unflavored $\mathcal{N}=4$ SYM on $\mathbb{R}^3\times \rm S^1$. For all %of them, we find \textit{inverse magnetic catalysis} consistent with the recent lattice QCD result \cite{Bali:2011qj}.

The outline of this paper is as follows: In section \ref{EM}, we write down the 5-dimensional Einstein-Maxwell action (which will be used to study confinement-deconfinement phase transition in hard-wall AdS/QCD and holographic dual of unflavored $\mathcal{N}=4$ SYM on $\mathbb{R}^3\times \rm S^1$) including the Gibbons-Hawking surface term and the appropriate counter terms. We then review the black hole solution in the presence of constant magnetic field $B\ll T^2$ found in \cite{D'Hoker:2009bc}. Then, starting from the black hole solution, which corresponds to the deconfined phase of strongly coupled QCD on flat spacetime and unflavored $\mathcal{N}=4$ SYM on $\mathbb{R}^3\times \rm S^1$, we derive the corresponding thermal-$AdS_{5}$ and $AdS_{5}$-soliton solutions, which correspond to the confined phases of strongly coupled QCD on flat spacetime and unflavored $\mathcal{N}=4$ SYM on $\mathbb{R}^3\times \rm S^1$, respectively.  We also determine the on-shell Euclidean actions (free energies) for the black hole, thermal-$AdS_{5}$, and $AdS_{5}$-soliton solutions.

In section \ref{hQCD}, we compute the critical temperature $T_{c}$ of the confinement-deconfinement phase transition in hard-wall AdS/QCD, and holographic duals of flavored and unflavored $\mathcal{N}=4$ SYM on $\mathbb{R}^3\times \rm S^1$. We first compute the critical temperature $T_{c}$ of the confinement-deconfinement phase transition in hard-wall AdS/QCD by requiring the difference between the black hole and thermal-$AdS_{5}$ on-shell Euclidean actions vanish at $T=T_{c}$. Then, we compute the critical temperature $T_{c}$ of the confinement-deconfinement phase transition in holographic dual of unflavored $\mathcal{N}=4$ SYM on $\mathbb{R}^3\times \rm S^1$ by requiring the difference between the black hole and $AdS_{5}$-soliton on-shell Euclidean actions vanish at $T=T_{c}$. Finally, using the insight we gained, in computing the critical temperature $T_{c}$ of the confinement-deconfinement phase transition in holographic dual of unflavored $\mathcal{N}=4$ SYM on $\mathbb{R}^3\times \rm S^1$, we compute the critical temperature $T_{c}$ of the confinement-deconfinement phase transition in holographic dual of flavored $\mathcal{N}=4$ SYM on $\mathbb{R}^3\times \rm S^1$ by constructing the backreacted $AdS_{5}$-soliton solution from the backreacted black hole metric of $\rm{D3/D7}$ model, with magnetic field $B$, found in \cite{Ammon:2012qs}.

In Appendix \ref{sec:large B}, we compute the critical temperature $T_{c}(B)$ of the confinement-deconfinement phase transition in hard-wall AdS/QCD for large magnetic field $B\gg T^2$.

\section{\label{EM}Einstein-Maxwell theory in 5D}
In this section, we review elements of Einstein-Maxwell theory in 5D which will, subsequently, be used to study confinement-deconfinement phase transitions in hard-wall AdS/QCD and holographic dual of unflavored $\mathcal{N}=4$ SYM on $\mathbb{R}^3\times \rm S^1$.

The action of five-dimensional Einstein-Maxwell theory with a negative cosmological constant is \cite{D'Hoker:2009bc}\footnote{Our conventions here are such that the Ricci scalar here $R_{here}$ is related to the Ricci scalar there $R_{there}$ (given in \cite{D'Hoker:2009bc}) by $R_{here}=-R_{there}$.} %the same as []:  %$R^\lambda_{~\mu\nu\kappa}= \p_\kappa \Gamma^\lambda_{\mu\nu}-\p_\nu \Gamma^\lambda_{\mu\kappa} %+\Gamma^\eta_{\mu\nu}\Gamma^\lambda_{\kappa\eta}-\Gamma^\eta_{\mu\kappa}\Gamma^\lambda_{\nu\eta}$
%and $R_{\mu\nu} = R^\lambda_{~\mu \lambda \nu}$.  }
\begin{equation}\label{action}
S= S_{\rm{bulk}}+S_{\rm{bndy}}~,
\end{equation}
where the bulk action $S_{\rm{bulk}}$ is
%\bea
\begin{equation}\label{bulkaction}
  S_{\rm{bulk}} = {1\over 16\pi G_5}\int\! d^5 x\sqrt{-g}\Big( R - F^{MN}F_{MN}+{12\over L^2} \Big)~,
\end{equation}
and the boundary action $S_{\rm{bndy}}$ is
\begin{equation}\label{boundaryaction}
S_{{\rm bndy}}={1\over 8\pi G_5} \int \! d^4x \sqrt{-\gamma}
\bigg( K -{3\over L}+
{L \over 2}  \left ( \ln {r \over L} \right ) F^{\mu\nu}F_{\mu\nu} \bigg )\bigg |_{r=r_{\Lambda}}~.
\end{equation}
%\eea
In the boundary action $S_{{\rm bndy}}$ (\ref{boundaryaction}), the first term is the Gibbons-Hawking surface term, and the other terms are the counter terms needed to cancel the UV($r_{\Lambda}\rightarrow\infty$) divergences in the bulk action in accordance with the holographic renormalization procedure \cite{Skenderis:2002wp}. Note that the counter terms are entirely constructed from the induced metric $\gamma_{\mu\nu}$ on the boundary surface at $r=r_{\Lambda}$, that is,
 \begin{equation}\label{inducedmetric}
   \gamma_{\mu \nu}(r_{\Lambda})= \text{diag}\left(g_{tt}(r_{\Lambda}),g_{xx}(r_{\Lambda}),g_{yy}(r_{\Lambda}),g_{zz}(r_{\Lambda})\right).
 \end{equation}
And, $K$ is the trace, with respect to $\gamma_{\mu\nu}$, of the extrinsic curvature
of the boundary given by $K_{\mu\nu} = (\p_r \gamma_{\mu\nu})/(2\sqrt{g_{rr}})$. Using the matrix formula $\partial_{\mu}(\det M)= \det M \textrm{tr}(M^{-1}\partial_{\mu}M)$ \cite{Natsuume:2014sfa}, we can write $K=\gamma^{\mu\nu}K_{\mu\nu}=\frac{\sqrt{g^{rr}}\p_r \sqrt{\gamma}}{\sqrt{\gamma}}$ \cite{D'Hoker:2009bc,Natsuume:2014sfa}.

In addition to the Bianchi identity, the field equations are \cite{D'Hoker:2009bc}
%The non-diffeomorphism invariant
%$\ln r$ term in the boundary  action is needed to remove the divergence associated
%with the trace anomaly $T^\mu_\mu \sim  F^{\mu\nu}F_{\mu\nu}$. Along with the Bianchi identity, the field equations are
%
%\bea\label{bb}
\begin{equation}\label{einstein}
R_{MN} = -{ 4\over L^2} g_{MN} -{1 \over 3} F^{PQ}F_{PQ} g_{MN} +2 F_{MP}F_N^{~P}~,
\end{equation}
\begin{equation}\label{maxwell}
\nabla^M F_{MN}  =  0~.
\end{equation}
%\eea
From now on we set the AdS radius to unity, that is, $L=1$.

Turning on a constant bulk magnetic field, in the $z$-direction, $B_{z}=F_{xy}=\partial_{x}A_{y}-\partial_{y}A_{x}=B$, where the bulk gauge potential $A_{\mu}(x,r)=\frac{1}{2}B(x\delta_{\mu}^{y}-y\delta_{\mu}^{x})$,\footnote{Note that the bulk gauge potential $A_{\mu}(x,r)$ and the corresponding bulk magnetic field $B=F_{xy}(x,r)$ are dual to the boundary gauge potential $A_{\mu}(x)=A_{\mu}(x,r=\infty)$ and the corresponding boundary magnetic field $B=F_{xy}(x,r=\infty)$ which couple to the $U(1)$ charged particles of the field theory living at the boundary. Later on, when we start discussing specific holographic models to QCD, we will specify the type of global $U(1)$ gauge group and the associated boundary current $J^{\mu}(x)$.} which solves Maxwell's equation (\ref{maxwell}), and contracting Einstein's field equation (\ref{einstein}), one can find the Ricci scalar $R=g^{MN}R_{MN}$ to be
\bea\label{ricci}
R&=& -20 + {2\over 3} B^2 g^{xx}g^{yy}.
\eea
So, the on-shell Euclidean action $S_{\rm{E}}$ (which can be found from the Lorentzian action (\ref{action}) by analytic continuation in the imaginary time direction, i.e., $t_{E}=it$) takes the form
%\begin{eqnarray}
\begin{equation}\label{oeaction}
S_{\rm{E}}=S_{\rm{bulk}}^{\rm{E}}+S^{\rm{E}}_{\rm{bndy}}~,
\end{equation}
where the on-shell Euclidean bulk action $S^{\rm{E}}_{\rm{bulk}}$ is
\begin{equation}\label{ebulkaction}
S_{\rm{bulk}}^{\rm{E}}={V_{3}\over 8\pi G_5} \int^{\beta}_{0}\! dt_{E}\int_{r'}^{r_{\Lambda}}\! dr\sqrt{g}\Big( 4 + \frac{2}{3}B^{2}g^{xx}g^{yy} \Big)~,
\end{equation}
and, the on-shell Euclidean boundary action $S^{\rm{E}}_{\rm{bndy}}$ is
%\bea
\begin{equation}\label{eboundaryaction}
S^{\rm{E}}_{{\rm bndy}}=-{V_{3}\over 8\pi G_5} \int^{\beta}_{0}\! dt_{E} \sqrt{\gamma}
\bigg( K -3 +B^2g^{xx}g^{yy}\ln {r_{\Lambda}} \bigg )~,
\end{equation}
%\eea
and, $r_{\Lambda}$ is the UV cut-off while $r'$ is the radius of the horizon $r'=r_{h}$ for a black hole solution, and IR cut-off $r'=r_{0}$ for a thermal-$AdS_{5}$ or $AdS_{5}$-soliton solutions. From now on we set $V_{3}=8\pi G_{5}=1$. Also, note that the on-shell Euclidean action $S_{\rm{E}}$ is related to the free energy $F$ by $S_{\rm{E}}=\beta F$.

\subsection{\label{Background}Background solutions with $B\ll T^2$}
In this subsection, we review the black hole solution in the presence of constant magnetic field $B\ll T^2$ found in \cite{D'Hoker:2009bc} which corresponds to the deconfined phase of strongly coupled QCD on $\mathbb{R}^4$ (flat spacetime) and unflavored $\mathcal{N}=4$ SYM on $\mathbb{R}^3\times \rm S^1$. Then, starting from the black hole solution, by setting the mass of the black hole to zero \cite{Lee:2009bya}, we derive the corresponding thermal-$AdS_{5}$ solution which is the holographic dual to the confined phase of strongly coupled QCD on flat spacetime $\mathbb{R}^4$. And, by "double Wick rotating" the black hole solution \cite{CasalderreySolana:2011us,Natsuume:2014sfa}, we derive the corresponding $AdS_{5}$-soliton solution which is the holographic dual to the confined phase of unflavored and strongly coupled $\mathcal{N}=4$ SYM on $\mathbb{R}^3\times \rm S^1$.

\subsubsection{Black hole}
For $B\ll T^{2}$ and electric charge density $\rho$, the perturbative black hole solution in powers of $B$, up to an integration constant $a_{3}$ is given in Eq. 6.1 and 6.16 of Ref. \cite{D'Hoker:2009bc}. Here, we set the electric charge density $\rho=0$ and fix the integration constant $a_{3}=-\frac{2}{3}$ so that the perturbative solution in powers of $B$ matches the near boundary solution which is also given in Eq. 4.4, 4.5 and 6.16 of \cite{D'Hoker:2009bc}. Therefore, the black hole solution in Eq. 6.1 and 6.16 of Ref. \cite{D'Hoker:2009bc}, for vanishing electric charge density $\rho=0$ and  $a_{3}=-\frac{2}{3}$, takes the form
\begin{eqnarray}\label{blackhole2}
ds_{\rm{bh}}^{2} &=& r^2 \left(-f(r) dt^2+q(r)dz^2+h(r)\left(dx^2+dy^2\right)\right)+\frac{dr^2}{f(r) r^2}~,
\\
f(r) &=& 1-\frac{M}{r^4}-\frac{2}{3}B^2\frac{\ln r}{r^4}+ \mathcal{O}(B^4)~,\nonumber
\\
q(r) &=& 1-\frac{2}{3}B^2\frac{\ln r}{r^4}+\mathcal{O}(B^4)~,\nonumber
\\
h(r) &=& 1+\frac{1}{3}B^2\frac{\ln r}{r^4}+\mathcal{O}(B^4)~, \nonumber
%\\
%T &=& \frac{1}{\beta} = U'(r_{h}) = \frac{r_h}{2\pi}\left(1+\frac{M}{r_{h}^4}+\frac{2}{3}B^2\left(\frac{\ln %r_{h}}{r_{h}^4}-\frac{1}{2r_{h}^4}\right)\right)+\mathcal{O}(B^4).
\end{eqnarray}
and, the Hawking temperature $T$ becomes
\begin{equation}\label{htemperature2}
 T = \frac{1}{\beta} = U'(r_{h}) = \frac{r_h}{2\pi}\left(1+\frac{M}{r_{h}^4}-\frac{2}{3}B^2\left(\frac{1}{2r_{h}^4}-\frac{\ln r_{h}}{r_{h}^4}\right)\right)+\mathcal{O}(B^4)~,
\end{equation}
where $M$ is the mass of the black hole, $U(r)=r^2f(r)$, the radius of the horizon $r_{h}$ is defined by requiring $f(r=r_{h})=0$, $T$ is the Hawking temperature of the black hole, and $\beta$ is the length of the thermal circle which acquired a fixed value as a function of $r_{h}$ in order to avoid the canonical singularity at the horizon $r=r_{h}$. One can also check that (\ref{blackhole2}) indeed satisfies the Einstein field equation (\ref{einstein}) or its contracted version (\ref{ricci}).

\subsubsection{Thermal-$\rm{AdS_{5}}$}
%\subsection{small $B$}
The thermal-$AdS_{5}$ solution can be found from a black hole solution by setting the mass of the black hole $M$ to zero, see \cite{Lee:2009bya} for the electrically charged black hole case. Therefore, from the black hole solution for $B\ll T^{2}$ (\ref{blackhole2}), we can determine the thermal-AdS solution for $B\ll \Lambda_{IR}^{2}\sim r_{0}^2$ by setting the mass of the black hole $M=0$,
\begin{eqnarray}\label{thermal}
ds_{\rm{thermal}}^{2} &=& r^2 \left(-f_{0}(r) dt^2+q(r)dz^2+h(r)\left(dx^2+dy^2\right)\right)+\frac{dr^2}{f_{0}(r) r^2}~,
\\
f_{0}(r) &=& 1-\frac{2}{3}B^2\frac{\ln r}{r^4}+ \mathcal{O}(B^4)~,\nonumber
\\
q(r) &=& 1-\frac{2}{3}B^2\frac{\ln r}{r^4}+\mathcal{O}(B^4)~,\nonumber
\\
h(r) &=& 1+\frac{1}{3}B^2\frac{\ln r}{r^4}+\mathcal{O}(B^4)~.\nonumber
\end{eqnarray}

\subsubsection{$\rm{AdS_{5}}$-soliton}
The $AdS_{5}$-soliton solution \cite{Horowitz:1998ha,Surya:2001vj} can be determined from the black hole solution (\ref{blackhole2}) by "double Wick rotation" $t=iz'$ and $z=it'$ \cite{CasalderreySolana:2011us,Natsuume:2014sfa}. Therefore, for $B\ll \Lambda_{IR}^{2}\sim r_{0}^2$ the $AdS_{5}$-soliton solution is,

\begin{eqnarray}\label{soliton}
ds_{\rm{soliton}}^{2} &=& r^2 \left(f_{s}(r)dz'^2-q(r)dt'^2+h(r)\left(dx^2+dy^2\right)\right)+\frac{dr^2}{f_{s}(r) r^2}~,
\\
f_{s}(r) &=& 1-\frac{M}{r^4}-\frac{2}{3}B^2\frac{\ln r}{r^4}+ \mathcal{O}(B^4)~,\nonumber
\\
q(r) &=& 1-\frac{2}{3}B^2\frac{\ln r}{r^4}+\mathcal{O}(B^4)~,\nonumber
\\
h(r) &=& 1+\frac{1}{3}B^2\frac{\ln r}{r^4}+\mathcal{O}(B^4)~, \nonumber
\\
\frac{1}{l} &=& \frac{U'(r_{0})}{4\pi} = \frac{r_0}{2\pi}\left(1+\frac{M}{r_{0}^4}+\frac{2}{3}B^2\left(\frac{\ln r_{0}}{r_{0}^4}-\frac{1}{2r_{0}^4}\right)\right)+\mathcal{O}(B^4). \nonumber
%\\
%&=& \frac{r_0}{\pi}\left(1-\frac{1}{6}\frac{B^2}{r_{0}^4}\right)+\mathcal{O}(B^4),
\end{eqnarray}
where $l$ is the length of the circle in the compactified $z'$ direction which is arbitrary for the black hole solution but in order to avoid the canonical singularity at $r=r_{0}$ (where $r_{0}$ is defined by requiring $f_{s}(r=r_{0})=0$), it acquires a finite value which is given in terms of $r_{0}$ for the $AdS_{5}$-soliton solution (\ref{soliton}).

\subsection{\label{On-shell}On-shell Euclidean actions with $B\ll T^2$}
In this subsection, we determine the on-shell Euclidean actions (free energies) for the black hole, thermal-$AdS_{5}$, and $AdS_{5}$-soliton solutions. And, we compute the difference between the on-shell Euclidean actions of the deconfining geometry (which is the black hole geometry for both hard-wall AdS/QCD and holographic dual of unflavored $\mathcal{N}=4$ SYM on $\mathbb{R}^3\times \rm S^1$) and the confining geometry (which is the thermal-$AdS_{5}$ geometry for hard-wall AdS/QCD, and the $AdS_{5}$-soliton geometry for holographic dual of unflavored $\mathcal{N}=4$ SYM on $\mathbb{R}^3\times \rm S^1$).

\subsubsection{Black hole}
The on-shell Euclidean action $S_{\rm{E}}=S_{\rm{bh}}$ (\ref{oeaction}) for the black hole solution with $B\ll T^{2}$ (\ref{blackhole2}) is
%$F_{bh}^{B\ll r_h^{2}}=\frac{S_{E}}{\beta}$ for $B\ll r_h^{2}$ is
\begin{eqnarray}\label{oebhaction}
S_{\rm{bh}} &=& S_{\rm{bulk}}+S_{\rm{bndy}},
\end{eqnarray}
where the on-shell Euclidean bulk action of the black hole $S_{\rm{bulk}}^{}$ for $B\ll T^{2}$ is
\begin{eqnarray}\label{oebhbulkaction}
S_{\rm{bulk}} &=& \int_{0}^{\beta}\! dt_{E}\int_{r_{h}}^{r_{\Lambda}}\! dr\sqrt{g}\Big( 4 + \frac{2}{3}B^{2}g^{xx}g^{yy} \Big),
\end{eqnarray}
and the on-shell Euclidean boundary action of the black hole $S_{\rm{bndy}}$ for $B\ll T^{2}$ is
\begin{equation}\label{oebhboundaryaction}
S_{\rm{bndy}}=-\int_{0}^{\beta}\! dt_{E}\sqrt{\gamma}\bigg(\frac{\sqrt{g^{rr}}\p_r \sqrt{\gamma}}{\sqrt{\gamma}} -3 +B^2g^{xx}g^{yy}\ln {r_{\Lambda}} \bigg ).
\end{equation}
The bulk action $S_{\rm{bulk}}$ (\ref{oebhbulkaction}) (after using the black hole metric for $B\ll T^{2}$ (\ref{blackhole2}), using the fact that $h(r)\sqrt{q(r)}=1+\mathcal{O}(B^4)$, evaluating the integrals, and simplifying) become %and taking the $r_{\Lambda}\rightarrow \infty$ %limit gives the finite result
\begin{eqnarray}\label{soebhbulkaction}
S_{\rm{bulk}} &=& -\beta\Big(r_{h}^4-r_{\Lambda}^4-\frac{2}{3}B^2\ln r_{\Lambda}+\frac{2}{3}B^2\ln r_{h}\Big)+\mathcal{O}(B^4),
\end{eqnarray}
which diverges when $r_{\Lambda}\rightarrow \infty$, and the boundary action $S_{\rm{bndy}}$ (\ref{oebhboundaryaction}) becomes
\begin{eqnarray}\label{soebhboundaryaction}
S_{\rm{bndy}}&=&-\beta(r_{\Lambda}^4+\frac{2}{3}B^2\ln r_{\Lambda}-\frac{1}{2}M-\frac{1}{3}B^2)+\mathcal{O}(B^4),
\end{eqnarray}
where we ignored terms which goes to zero in the $r_{\Lambda}\rightarrow \infty$ limit. Also note that (\ref{soebhboundaryaction}) diverges when $r_{\Lambda}\rightarrow \infty$, but the sum of $S_{\rm{bulk}}$ (\ref{soebhbulkaction}) and $S_{\rm{bndy}}$ (\ref{soebhboundaryaction}) is finite. Hence, the black hole on-shell Euclidean action $S_{\rm{bh}}$ (\ref{oebhaction}) is
\begin{eqnarray}\label{finalbh}
S_{\rm{bh}} &=& S_{\rm{bulk}}+S_{\rm{bndy}}=-\beta\left(r_{h}^4-\frac{1}{2}M+\frac{2}{3}B^2\ln r_{h}-\frac{1}{3}B^2\right)+\mathcal{O}(B^4)~.
\end{eqnarray}

%////////////////////////////////////////////

\subsubsection{Thermal-$AdS_{5}$}
The on-shell Euclidean action $S_{\rm{E}}=S_{\rm{thermal}}$ (\ref{oeaction}) for the thermal-$AdS_{5}$ solution with $B\ll \Lambda_{IR}^{2}\sim r_{0}^2$ (\ref{thermal}) is
\begin{eqnarray}\label{oetaction}
S_{\rm{thermal}} &=& S_{\rm{tbulk}}+S_{\rm{tbndy}},
\end{eqnarray}
where the on-shell Euclidean bulk action $S_{\rm{tbulk}}$ of the thermal-$AdS_{5}$ for $B\ll \Lambda_{IR}^{2}\sim r_{0}^2$  is
\begin{eqnarray}\label{oetbulkaction}
S_{\rm{tbulk}} &=& \int_{0}^{\beta'}\! dt_{E}\int_{r_{0}}^{r_{\Lambda}}\! dr\sqrt{g}\Big( 4 + \frac{2}{3}B^{2}g^{xx}g^{yy} \Big),
\end{eqnarray}
and the on-shell Euclidean boundary action $S_{\rm{tbndy}}$ of the thermal-$AdS_{5}$ for $B\ll T^{2}$ is
\begin{equation}\label{oetboundaryaction}
S_{\rm{tbndy}}=-\int_{0}^{\beta'}\! dt_{E}\sqrt{\gamma}\bigg(\frac{\sqrt{g^{rr}}\p_r \sqrt{\gamma}}{\sqrt{\gamma}} -3 +B^2g^{xx}g^{yy}\ln {r_{\Lambda}} \bigg ).
\end{equation}
The thermal-$AdS_{5}$ bulk action $S_{\rm{tbulk}}$ (\ref{oetbulkaction}) (after using the thermal-$AdS_{5}$ metric for $B\ll T^{2}$ (\ref{thermal}), using the fact that $h(r)\sqrt{q(r)}=1+\mathcal{O}(B^4)$, evaluating the integrals, and simplifying) becomes %and taking the $r_{\Lambda}\rightarrow \infty$ %limit gives the finite result
\begin{eqnarray}\label{soetbulkaction}
S_{\rm{tbulk}} &=& -\beta'\Big(r_{0}^4-r_{\Lambda}^4-\frac{2}{3}B^2\ln r_{\Lambda}+\frac{2}{3}B^2\ln r_{0}\Big)+\mathcal{O}(B^4),
\end{eqnarray}
which diverges when $r_{\Lambda}\rightarrow \infty$, and the thermal-$AdS_{5}$ boundary action $S_{\rm{tbndy}}$ (\ref{oetboundaryaction}) becomes
\begin{eqnarray}\label{soetboundaryaction}
S_{\rm{tbndy}}&=&-\beta'(r_{\Lambda}^4+\frac{2}{3}B^2\ln r_{\Lambda}-\frac{1}{3}B^2)+\mathcal{O}(B^4),
\end{eqnarray}
which diverges as well when $r_{\Lambda}\rightarrow \infty$. But, the sum of $S_{\rm{tbulk}}$ (\ref{soetbulkaction}) and $S_{\rm{tbndy}}$ (\ref{soetboundaryaction}) is finite. Hence, the thermal on-shell Euclidean action $S_{\rm{thermal}}$ (\ref{oetaction}) becomes
\begin{eqnarray}\label{finalt}
S_{\rm{thermal}} &=& -\beta(r_{0}^4+\frac{2}{3}B^2\ln r_{0}-\frac{1}{3}B^2)+\mathcal{O}(B^4).
\end{eqnarray}
where we used $\beta'=\beta\sqrt{f}=\beta$ at the boundary $r_{\Lambda}\rightarrow \infty$.

Therefore, $\Delta S_{\rm{E}}$ (which is the difference between the $AdS_{5}$ black hole (\ref{finalbh}) and thermal-$AdS_{5}$ (\ref{finalt}) on-shell Euclidean actions) becomes
\begin{eqnarray}\label{difference}
\Delta S_{\rm{E}} &=& S_{\rm{bh}}-S_{\rm{thermal}}=-\beta\left(r_{h}^4-r_{0}^4-\frac{1}{2}M+\frac{2}{3}B^2\ln(\frac{r_{h}}{r_{0}})\right)+\mathcal{O}(B^4)~.
\end{eqnarray}

\subsubsection{$\rm{AdS_{5}}$-soliton}
Since, black hole (\ref{blackhole2}) and $AdS_{5}$-soliton (\ref{soliton}) are equivalent Euclidean geometries, their on-shell Euclidean actions take similar form. In fact, the on-shell Euclidean action of  $AdS_{5}$-soliton can be found by merely replacing $r_{h}$ by $r_{0}$ in the on-shell Euclidean action for the black hole \cite{Natsuume:2014sfa}.  Therefore, the difference between the on-shell actions $S_{\rm{bh}}$ of the black hole (\ref{finalbh}) and $S_{\rm{soliton}}$ of $AdS_{5}$-soliton geometries is simply
\begin{eqnarray}\label{deltasoliton}
\Delta S_{\rm{E}}&=& S_{\rm{bh}}-S_{\rm{soliton}}=-\beta\left(r_{h}^4-r_{0}^4+\frac{2}{3}B^2\ln \frac{r_{h}}{r_{0}}\right)+\mathcal{O}(B^4)~.
\end{eqnarray}

\section{\label{hQCD}Confinement-deconfinement phase transition in holographic models of QCD for $B\ll T^2$}
In this section, we compute the critical temperature $T_{c}$ of the confinement-deconfinement phase transition in hard-wall AdS/QCD, and holographic duals of flavored and unflavored $\mathcal{N}=4$ SYM on $\mathbb{R}^3\times \rm S^1$.

\subsection{\label{hard}Confinement-deconfinement phase transition in hard-wall AdS/QCD}
\begin{figure}[t]
	\centering
	\includegraphics[scale=0.6]{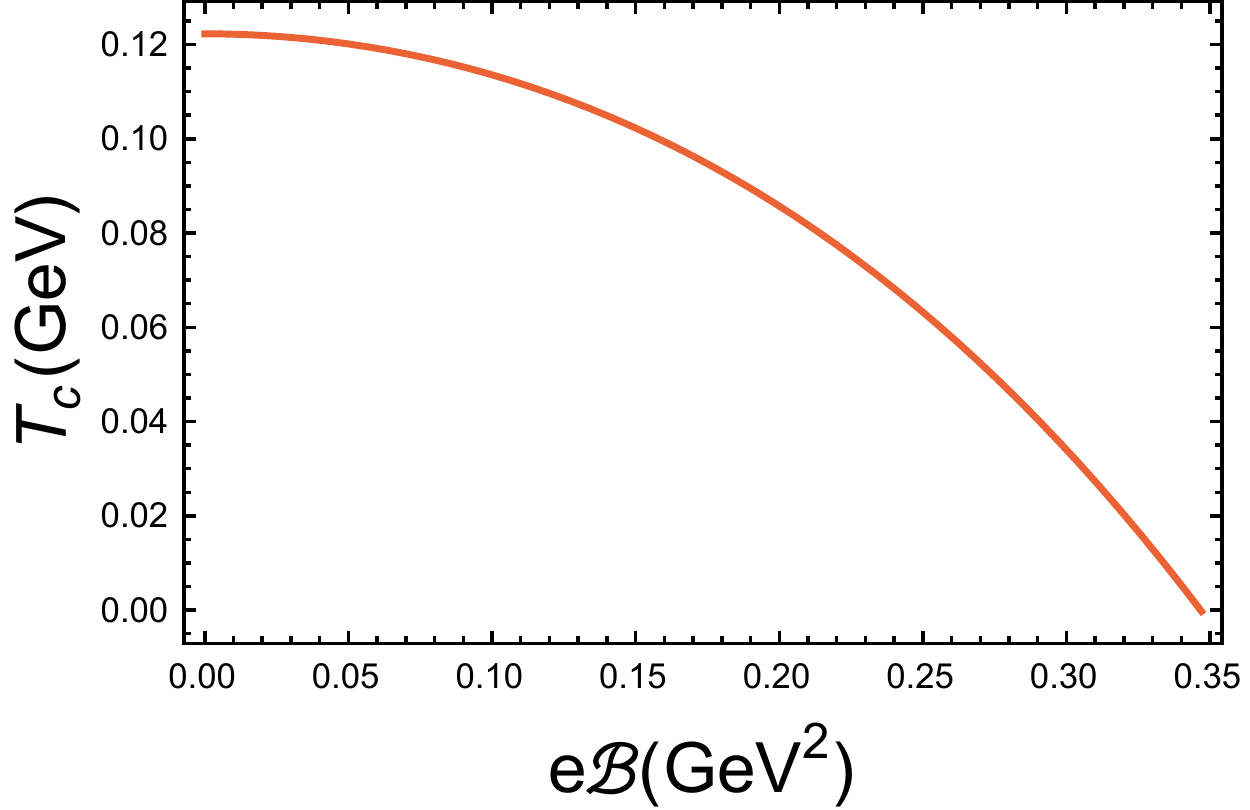}
		\caption{Critical temperature $T_{c}(B)$ of the hard-wall AdS/QCD with $r_{0}=\frac{m_{\rho}}{2.405}=0.323~GeV$ \cite{Herzog:2006ra}. Note: $\mathcal{B}=\sqrt{3}B$ is the physical magnetic field at the boundary \cite{D'Hoker:2009mm}. \label{fig1}}
\end{figure}

For hard-wall AdS/QCD\cite{D'Hoker:2009bc}\footnote{For the hard-wall AdS/QCD model, the bulk magnetic field $B=F_{xy}(x,r)$ and the corresponding bulk gauge potential $A_{\mu}(x,r)=\frac{1}{2}B(x\delta_{\mu}^{y}-y\delta_{\mu}^{x})$ are dual to the boundary magnetic field $B=F_{xy}(x,r=\infty)$ and the corresponding boundary gauge potential $A_{\mu}(x)=A_{\mu}(x,r=\infty)$ of the $U(1)_{V}$ subgroup of the $SU(N_{f})$ global flavor group of QCD. And, the boundary vector gauge potential $A_{\mu}(x)$ couples to the boundary conserved vector current $J^{\mu}_{V}(x)$.}, we determine the critical temperature $T_{c}(B)$ of the confinement-deconfinement phase transition by first determining the critical radius of the horizon $r_{h}=r_{hc}$ from the condition that the difference between the Euclidean actions for the black hole and thermal-$AdS_{5}$ solutions vanish at $r_{h}=r_{hc}$, i.e., $\Delta S_{\rm{E}}(r_{h}=r_{hc})=0$. For $B\ll T^{2}$, requiring $\Delta S_{\rm{E}}(r_{hc})=0$ in (\ref{difference}), we find the constraint equation for the critical radius of the horizon $r_{hc}$ to be
\begin{equation}\label{constraint}
r_{hc}^4+\frac{2}{3}B^2\ln(\frac{r_{hc}}{r_{0}})-2r_{0}^4+\mathcal{O}(B^4)=0,
\end{equation}
which can be solved numerically for $r_{hc}(B,r_{0})$. Note that, we have fixed $M=2r_{0}^4$ in (\ref{difference}), so that (\ref{constraint}) reduces to the constraint equation found in \cite{Herzog:2006ra,BallonBayona:2007vp} at $B=0$, which is $r_{hc}^4=2r_{0}^4$. Once we find the solution for $r_{hc}$ from the constraint equation (\ref{constraint}), we can use (\ref{htemperature2}) to find $T_{c}=T(r_{h}=r_{hc}, M=2r_{0}^4)$. The plot of the numerical solution for $T_{c}(B,r_{0})$ for $B\ll T^{2}$ is given in Fig.~\ref{fig1}, and the numerical plot clearly shows that $T_{c}(B)$ decreases with increasing $B\ll T^2$ in agreement with the \textit{inverse magnetic catalysis} recently found in lattice QCD for $B\lesssim1~GeV^2$ \cite{Bali:2011qj}.

\subsection{\label{unflavored}Confinement-deconfinement phase transition in holographic dual of unflavored $\mathcal{N}=4$ SYM on $\mathbb{R}^3\times \rm S^1$}
For the holographic dual of unflavored $\mathcal{N}=4$ SYM on $\mathbb{R}^3\times \rm S^1$\footnote{For the unflavored $\mathcal{N}=4$ SYM case, the bulk magnetic field $B=F_{xy}(x,r)$ and the corresponding bulk gauge potential $A_{\mu}(x,r)=\frac{1}{2}B(x\delta_{\mu}^{y}-y\delta_{\mu}^{x})$ are dual to the boundary magnetic field $B=F_{xy}(x,r=\infty)$ and the boundary gauge potential $A_{\mu}(x)=A_{\mu}(x,r=\infty)$ of the $U(1)$ subgroup of the $SU(4)_{R}$ global R-symmetry group of $\mathcal{N}=4$ SYM which couples to the boundary R-current $J_{R}^{\mu}$.}, we study the confinement-deconfinement phase transition by using the same Einstein-Maxwell action in $5D$ as we used for the hard-wall AdS/QCD, and the analysis will be similar to the hard-wall AdS/QCD case but, for the unflavored $\mathcal{N}=4$ SYM on $\mathbb{R}^3\times \rm S^1$ case, we compactify the black hole solution in the $z$ direction into a circle of length $l$, and compare its free energy with the free energy of $AdS_{5}$-soliton solution (\ref{soliton}) instead of the thermal-$AdS_{5}$ solution (\ref{thermal}) that we used for the hard-wall AdS/QCD.

It is easy to see from (\ref{deltasoliton}) that the critical radius of the horizon $r_{h}=r_{hc}$ at which $\Delta S_{\rm{E}}(r_{h}=r_{hc})=0$ is given by $r_{h}=r_{hc}=r_{0}$. Therefore, using (\ref{htemperature2}), the critical temperature $T_{c}=T(r_{h}=r_{hc}=r_{0})$ becomes,
\begin{equation}\label{critical2}
 T_{c} = \frac{r_0}{2\pi}\left(1+\frac{M}{r_{0}^4}-\frac{2}{3}B^2\left(\frac{1}{2r_{0}^4}-\frac{\ln r_{0}}{r_{0}^4}\right)\right)+\mathcal{O}(B^4)=\frac{1}{l}~.
\end{equation}
Fixing $M=r_{0}^4$ so that we reproduce the correct $B=0$ result $T_{c}(B=0)=\frac{r_{0}}{\pi}$, and fixing $r_{0}$ from the value of $T_{c}$ at $B=0$, which we denote as $T_{c}^0$, we can write (\ref{critical2}) in terms of $T_{c}^0=\frac{r_{0}}{\pi}$ as
\begin{equation}\label{ufcritical}
 T_{c} = T_{c}^0\left(1-\left(\frac{B}{B_{c}}\right)^2\right)+\mathcal{O}(B^4)~
\end{equation}
where we defined the critical magnetic field $B_{c}=\frac{\sqrt{6}\pi^2(T_{c}^0)^2}{1-2\ln (L T_{c}^0\pi)}$ and $L$ is the radius of the AdS spacetime. From (\ref{ufcritical}), it is easy to see that $T_{c}$ is a decreasing function with increasing $B\ll T^2$ in qualitative agreement with the recent lattice QCD result \cite{Bali:2011qj}.

\subsection{\label{flavored}Confinement-deconfinement phase transition in holographic dual of flavored $\mathcal{N}=4$ SYM on $\mathbb{R}^3\times \rm S^1$}

In the previous subsection, we have studied the confinement-deconfinement phase transition in the holographic dual of unflavored $\mathcal{N}=4$ SYM on $\mathbb{R}^3\times \rm S^1$ using the backreacted black hole and $AdS_{5}$-soliton geometries, from which, we can infer a simple prescription of finding $T_{c}$ in any backreacted black hole and $AdS_{5}$-soliton based models.

The prescription is, first find the backreacted metric and the Hawking temperature $T(r_{h})$ of the black hole, then the critical temperature $T_{c}$ is simply given by $T_{c}=T(r_{h}=r_{0})$ where $r_{0}$ can be fixed by the value of $T_{c}^0=T_{c}(B=0)$.

Therefore, using this prescription, we can determine $T_{c}(N_{f},B)$ of the holographic dual of flavored $\mathcal{N}=4$ SYM on $\mathbb{R}^3\times \rm S^1$. To this end we will use the backreacted metric of $\rm{D3/D7}$ model given in \cite{Ammon:2012qs} where the authors have also found the Hawking temperature $T(\frac{N_{f}}{N_{c}},B)$ including the backreaction of $N_{f}~D7$-branes and magnetic field $B$\footnote{For the $D3/D7$ model, we use a bulk magnetic field which corresponds to the Kalb-Ramond two form field $B_{xy}(x,r)$. And, also note that for the DBI action of the probe $D7$ brane, the gauge invariant and physically significant field strength is given by $\mathcal{F}_{mn}=F_{mn}+B_{mn}=\partial_{m} \mathcal{A}_{n}(x)-\partial_{n}\mathcal{A}_{m}(x)$ which is the sum of Maxwell's field strength $F_{mn}$ and the Kalb-Ramond two form field $B_{mn}$. And, the bulk magnetic field $B_{xy}(x,r)$ and the corresponding bulk gauge potential $\mathcal{A_{\mu}}(x,r)$ are dual to the boundary magnetic field $B=B_{xy}(x,r=\infty)$ and the boundary gauge potential $\mathcal{A_{\mu}}(x)=\mathcal{A_{\mu}}(x,r=\infty)$ of the $U(1)_{V}$ subgroup of the $SU(N_{f})$ global flavor group of the $\mathcal{N}=2$ supersymmetric field theory which couples to the boundary vector current $\mathcal{J}_{V}^{\mu}(x)$, see, for example, Ref. \cite{Hoyos:2011us}. For our case, $F_{mn}=0$ but $B_{mn}=B_{xy}$.} to be, see Eq. 3.1 of \cite{Ammon:2012qs}\footnote{In Eq. 3.1 of \cite{Ammon:2012qs}, $T$ is written in terms of $r_{m}^4$ and $\epsilon_{h}$. Here, we have used Eq. 2.35 and 3.4 of \cite{Ammon:2012qs} (which relates $\epsilon_{h}=\frac{\lambda_{h}}{8\pi^2}\frac{N_{f}}{N_{c}}$ and $r_{m}^2=B$, respectively) in order to write $T$ explicitly in terms of $\lambda_{h}$, $\frac{N_{f}}{N_{c}}$, and $B$.},
\begin{equation}\label{htf}
  T=\frac{r_{h}}{\pi}\left(1+\frac{\lambda_{h}}{64\pi^2}\frac{N_{f}}{N_{c}}\left(1-2\sqrt{1+\frac{B^2}{r_{h}^4}}\right)\right) + \mathcal{O}((N_{f}/N_{c})^2).
\end{equation}
Since the on-shell Euclidean action of the black hole solution (including the backreaction of $N_{f}~D7$-branes and magnetic field $B$) has also been given in Eq. 3.14 of \cite{Ammon:2012qs}, in order to find the corresponding Euclidean action of the $AdS_{5}$-soliton, all we need to do is replace $r_{h}$ by $r_{0}$ in Eq. 3.14 of \cite{Ammon:2012qs}. Hence, the difference between the two on-shell Euclidean actions vanishes at the critical radius of the horizon $r_{h}=r_{hc}=r_{0}$. And, using $r_{h}=r_{hc}=r_{0}$ in (\ref{htf}), the critical temperature $T_{c}=T(r_{h}=r_{hc}=r_{0})$ of the confinement-deconfinement phase transition in flavored $\mathcal{N}=4$ SYM on $\mathbb{R}^3\times \rm S^1$ becomes
\begin{equation}\label{}
  T_{c}=\frac{r_{0}}{\pi}\left(1+\frac{\lambda_{h}}{64\pi^2}\frac{N_{f}}{N_{c}}\left(1-2\sqrt{1+\frac{B^2}{r_{0}^4}}\right)\right) + \mathcal{O}((N_{f}/N_{c})^2),
\end{equation}
which can be written in terms of $T_{c}^0=T_{c}(N_{f}=0,B=0)=\frac{r_{0}}{\pi}$ as
\begin{equation}\label{fcritical}
  T_{c}=T_{c}^0\left(1+\frac{\lambda_{h}}{64\pi^2}\frac{N_{f}}{N_{c}}\left(1-2\sqrt{1+\frac{1}{\pi^4}\frac{B^2}{(T_{c}^0)^4}}\right)\right) + \mathcal{O}((N_{f}/N_{c})^2),
\end{equation}
where $\lambda_{h}$ is the value of the 't Hooft coupling fixed at the horizon $r_{h}$, that is, $\lambda_{h}=4\pi g_{s}e^{\phi_{h}}N_{c}$ where $g_{s}$ is the string coupling constant and $\phi(r)$ is the dilaton scalar field.

%Note that (\ref{fcritical}) always decreases with increasing $B$ as long as $\lambda_{h}>0$ and $\frac{N_{f}}{N_{c}}>0$, and vanishes at some critical %magnetic field $B_{c}$ given by
% \begin{equation}\label{}
%  B_{c}=(\pi T_{c}^0)^2\sqrt{\frac{32\pi^2}{\lambda_{h}}\frac{N_{c}}{N_{f}}+\left(\frac{32\pi^2}{\lambda_{h}}\frac{N_{c}}{N_{f}}\right)^2-\frac{3}{4}}.
%\end{equation}
Note that, for $B=0$, (\ref{fcritical}) reduces to
\begin{equation}\label{}
  T_{c}=T_{c}^0\left(1-\frac{\lambda_{h}}{64\pi^2}\frac{N_{f}}{N_{c}}\right) + \mathcal{O}((N_{f}/N_{c})^2),
\end{equation}
which is in a qualitative agreement with the hard-wall AdS/QCD \cite{Kim:2007em}, functional renormalization group study of QCD \cite{Braun:2006jd}, and lattice QCD \cite{Karsch:2000kv} results which show that $T_{c}$ decreases with increasing number of flavors $N_{f}$ at zero magnetic field $B=0$ and chemical potential $\mu=0$.

\begin{figure}[t]
	\centering
	\includegraphics[scale=0.6]{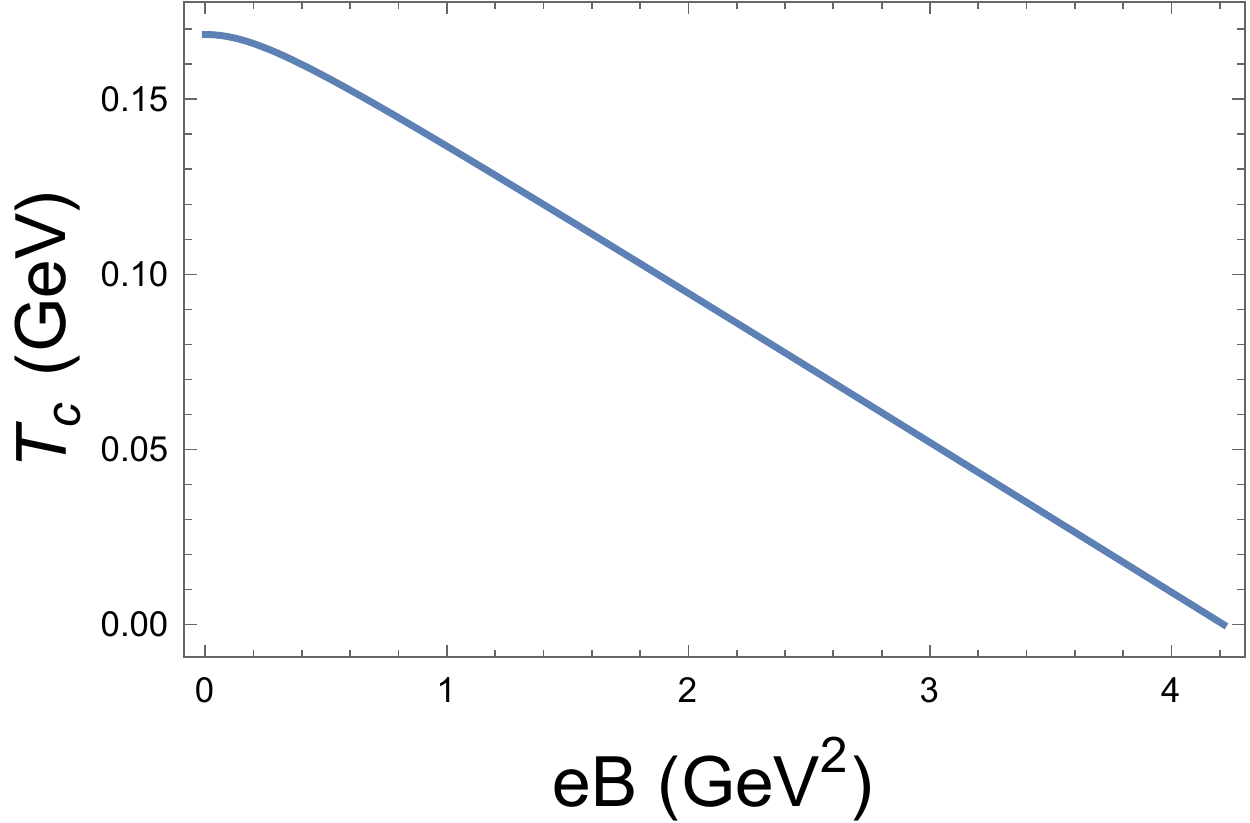}
		\caption{Critical temperature $T_{c}(B)$ of flavored $\mathcal{N}=4$ SYM on $R^3\times S^1$ (\ref{fcritical}) using $T_{c}^0=0.175~GeV$ and $\lambda_{h}=23\times\frac{N_{c}}{N_{f}}$. \label{fig2}}
\end{figure}

We have plotted (\ref{fcritical}) in Fig.~\ref{fig2} which clearly shows that $T_{c}(B)$ decreases with increasing $B\ll T^2$ in agreement with the \textit{inverse magnetic catalysis} recently found in lattice QCD for $B\lesssim1~GeV^2$ \cite{Bali:2011qj}.
%\begin{figure}[t]
%	\centering
%	\includegraphics[scale=0.6]{lTc2.pdf}
%		\caption{Critical temperature $T_{c}(B)$ of flavored $\mathcal{N}=4$ SYM on $R^3\times S^1$ (\ref{fcritical}) using $T_{c}^0=0.175~GeV$ and %$\lambda_{h}=23\times\frac{N_{c}}{N_{f}}$. \label{fig2}}
%\end{figure}

\section{Conclusion}
The observation of \textit{inverse magnetic catalysis}, for $B\lesssim1~GeV^2$, in the recent study of QCD on the lattice \cite{Bali:2011qj}, instead of \textit{magnetic catalysis} as predicted by most models of QCD \cite{Gusynin:1994re,Miransky:2002rp,Mizher:2010zb,Fraga:2008um,Gatto:2010pt,Gatto:2010qs,Osipov:2007je,Kashiwa:2011js,Klimenko:1992ch,Alexandre:2000yf,Filev:2007gb,Albash:2007bk,Alam:2012fw,Johnson:2008vna,Bergman:2008sg,Evans:2010xs}, has posed a serious challenge both for the holographic and non-holographic models of QCD. Therefore, motivated by gaining new insight into this problem, we have studied the $T-B$ phase diagrams of hard-wall AdS/QCD, and holographic duals of flavored and unflavored $\mathcal{N}$=4 SYM on $\mathbb{R}^3\times \rm S^1$ for $B\ll T^2$.

We have found that the $T-B$ phase diagrams of hard-wall AdS/QCD, and holographic duals of flavored and unflavored $\mathcal{N}$=4 SYM on $\mathbb{R}^3\times \rm S^1$ (which are based on the study of the confinement-deconfinement phase transition) are consistent with the recent lattice QCD result \cite{Bali:2011qj} but opposite to the results of the other holographic models of QCD (which are based on the study of the chiral-symmetry breaking phase transition) \cite{Filev:2007gb,Albash:2007bk,Alam:2012fw,Johnson:2008vna,Bergman:2008sg,Evans:2010xs}. As it can be seen in Fig.~\ref{fig1}, Fig.~\ref{fig2}, and Eq.(\ref{ufcritical}), we have found that the critical temperature $T_{c}(B)$ decreases with increasing $B\ll T^2$ in agreement with the \textit{inverse magnetic catalysis} observed in the recent lattice QCD result for $B\lesssim1~GeV^2$\cite{Bali:2011qj}.

However, we would like to emphasize the fact that the vanishing of $T_{c}(B)$, at some critical magnetic field $B_{c}$, as it can be seen in Fig.~\ref{fig1}, Fig.~\ref{fig2}, and Eq.(\ref{ufcritical}), might be the artifact of our usage of the backreacted black hole solution only for small magnetic field $B\ll T^2$ and $N_{f}\ll N_{c}$ limits. Had we used the exact solution for the backreacted black hole solution, for any value of $B$ and $N_{f}$, the critical temperature $T_{c}(B)$ might not necessarily vanish at some critical magnetic field $B_{c}$.

For example, in the large magnetic field $B\gg T^2$ limit, the Hawking temperature of the black hole geometry of \cite{D'Hoker:2009mm} is given by $T(B\gg T^2)=\frac{3}{2}\frac{r_{h}}{\pi}$ (\ref{htemperature222}), which according to the prescription of finding the critical temperature $T_{c}$ given in subsection \ref{flavored}, the critical temperature $T_{c}$ of the unflavored $\mathcal{N}=4$ SYM on $\mathbb{R}^3\times \rm S^1$ for large magnetic field $B\gg T^2$ will be $T_{c}(B\gg T^2)=\frac{3}{2}\frac{r_{0}}{\pi}=\frac{3}{2}T_{c}(B=0)$, which means that $T_{c}(B\gg T^2)>T_{c}(B=0)$. Therefore, $T_{c}(B)$, of unflavored $\mathcal{N}=4$ SYM on $\mathbb{R}^3\times \rm S^1$, is not a monotonically decreasing function of $B$, and it will eventually start to increase for large magnetic field $B$ and asymptotes to the constant value $T_{c}(B\gg T^2)=\frac{3}{2}T_{c}(B=0)$.

Similarly, for hard-wall AdS/QCD, the critical temperature $T_{c}(B\gg T^2)=\frac{3}{\sqrt{2}}\frac{r_{0}}{\pi}$ (\ref{largebtc}), but from (\ref{constraint}), one can see that $T_{c}(B=0)=2^{1/4}\frac{r_{0}}{\pi}$, see also \cite{Herzog:2006ra,BallonBayona:2007vp}, which means that $T_{c}(B\gg T^2)>T_{c}(B=0)$. Therefore, $T_{c}(B)$ of hard-wall AdS/QCD, is not a monotonically decreasing function of $B$, and it will eventually start to increase for large magnetic field $B$ and asymptotes to the constant value $T_{c}(B\gg T^2)=\frac{3}{\sqrt{2}}\frac{r_{0}}{\pi}$. We leave the holographic computation of $T_{c}(B)$ for arbitrary value of $B$ for future investigations.

This eventual increasing (after initially decreasing for small $B$) of $T_{c}(B)$ for large $B$ has also been predicted in some very recent non-holographic studies of the chiral-symmetry breaking phase transition in QCD \cite{Ferreira:2014kpa,Braun:2014fua}, and it has been related to the asymptotic freedom and dimensional reduction of QCD at large magnetic field $B$.

The \textit{inverse magnetic catalysis} we found in this paper, from our study of the confinement-deconfinement phase transition in the holographic models of QCD (hard-wall AdS/QCD, and holographic duals of flavored and unflavored $\mathcal{N}$=4 SYM on $\mathbb{R}^3\times \rm S^1$), is also qualitatively similar to the inverse magnetic catalysis observed in the non-holographic studies of the confinement-deconfinement phase transition in QCD, such as, MIT bag model \cite{Fraga:2012fs}, chiral perturbation theory with two quark flavors \cite{Agasian:2008tb}, and QCD at large-$N_{c}$ \cite{Fraga:2012ev}.

\acknowledgments
The author thanks Ho-Ung Yee for stimulating discussions, critical reading of the draft, and giving me important feedbacks which greatly improved the clarity of the draft. The author also thanks Bo Ling, and Misha Stephanov for very helpful discussions.

\appendix

\section{\label{sec:large B} Confinement-deconfinement phase transition in hard-wall AdS/QCD for $B\gg T^2$}
In this appendix, we compute the critical temperature $T_{c}(B)$ of the confinement-deconfinement phase transition in hard-wall AdS/QCD for large magnetic field $B\gg T^2$.

For $B\gg T^{2}$, the black hole solution is $AdS_{3}\times T^2$ black hole or $BTZ\times T^2$\cite{D'Hoker:2009mm}
\begin{eqnarray}\label{blackhole222}
ds_{\rm{bh}}^2 &=& 3r^2 \left(-f(r) dt^2+dz^2\right)+\frac{B}{\sqrt{3}}\left(dx^2+dy^2\right)+\frac{dr^2}{3f(r) r^2}~,
\\
f(r) &=& 1-\frac{M}{r^2}~.
\end{eqnarray}
and, the Hawking temperature $T$ becomes
\begin{equation}\label{htemperature222}
 T = \frac{1}{\beta} = U'(r_{h}) =\frac{3}{2}\frac{r_h}{\pi}~,
\end{equation}
where $M=r_{h}^2$ is the mass of the black hole, $U(r)=r^2f(r)$, the radius of the horizon $r_{h}$ is defined by requiring $f(r=r_{h})=0$, and $\beta$ is the length of the thermal circle. 

Note that the transverse $x-y$ plane to the direction of the magnetic field $B=B_{z}$ of the black hole solution (\ref{blackhole222}) is compactified to a 2-torus $T^2$ \cite{D'Hoker:2009mm} which is similar to the dimensional reduction that appears in field theories in an external magnetic field $B$.

For $B\gg \Lambda_{IR}^{2}\sim r_{0}^2$, the thermal-$AdS_{3}\times T^2$ solution can be found from the black hole solution by setting the mass of the black hole $M$ to zero, that is,
\begin{eqnarray}\label{thermalll}
ds_{\rm{thermal}}^2 &=& 3r^2 \left(-dt^2+dz^2\right)+\frac{B}{\sqrt{3}}\left(dx^2+dy^2\right)+\frac{dr^2}{3r^2}~.
\end{eqnarray}

The on-shell Euclidean action $S_{\rm{E}}=S_{\rm{bh}}$ (\ref{oeaction}) for the black hole solution with $B\gg T^{2}$ (\ref{blackhole222}) is
%$F_{bh}^{B\ll r_h^{2}}=\frac{S_{E}}{\beta}$ for $B\ll r_h^{2}$ is
\begin{eqnarray}\label{oebhactionnn}
S_{\rm{bh}} &=& S_{\rm{bulk}}+S_{\rm{bndy}},
\end{eqnarray}
where the on-shell Euclidean bulk action $S_{\rm{bulk}}^{}$ of the black hole solution for $B\gg T^{2}$ is
\begin{eqnarray}\label{oebhbulkactionnn}
S_{\rm{bulk}} &=& \int_{0}^{\beta}\! dt_{E}\int_{r_{h}}^{r_{\Lambda}}\! dr\sqrt{g}\Big( 4 + \frac{2}{3}B^{2}g^{xx}g^{yy} \Big),
\\
&=& 6\int_{0}^{\beta}\! dt_{E}\int_{r_{h}}^{r_{\Lambda}}\! dr\sqrt{g}~,
\end{eqnarray}
and, it turned out, we will not need the on-shell Euclidean boundary action of the black hole to remove the UV divergences, so we set $S_{\rm{bndy}}=0$ for $B\gg T^{2}$. Hence, the on-shell Euclidean action of the black hole $S_{\rm{bh}}$ (\ref{oebhactionnn}) becomes
\begin{eqnarray}\label{}
S_{\rm{bh}} &=& S_{\rm{bulk}}+S_{\rm{bndy}}=6\int_{0}^{\beta}\! dt_{E}\int_{r_{h}}^{r_{\Lambda}}\! dr\sqrt{g}~,
\end{eqnarray}
which (after using the black hole metric for $B\gg T^{2}$ (\ref{blackhole222}), evaluating the integrals, and simplifying) gives
\begin{eqnarray}\label{finalbhhh}
S_{\rm{bh}} &=& 3\beta B\left(r_{\Lambda}^2-r_{h}^2\right).
\end{eqnarray}

The on-shell Euclidean action $S_{\rm{E}}=S_{\rm{thermal}}$ (\ref{oeaction}) for the thermal-$AdS$ solution with $B\gg \Lambda_{IR}^{2}\sim r_{0}^2$ (\ref{thermalll}) is
\begin{eqnarray}\label{oetactionnn}
S_{\rm{thermal}} &=& S_{\rm{tbulk}}+S_{\rm{tbndy}},
\end{eqnarray}
where the on-shell Euclidean bulk action $S_{\rm{tbulk}}$ of the thermal-$AdS$ for $B\gg \Lambda_{IR}^{2}\sim r_{0}^2$  is
\begin{eqnarray}\label{oetbulkactionnn}
S_{\rm{tbulk}} &=& \int_{0}^{\beta'}\! dt_{E}\int_{r_{0}}^{r_{\Lambda}}\! dr\sqrt{g}\Big( 4 + \frac{2}{3}B^{2}g^{xx}g^{yy} \Big),
\end{eqnarray}
and, it turned out, we will not need the on-shell Euclidean boundary action $S_{\rm{tbndy}}$ of the thermal-$AdS$ for $B\gg \Lambda_{IR}^{2}\sim r_{0}^2$ to remove the UV divergences, so we set $S_{\rm{tbndy}}=0$ for $B\gg \Lambda_{IR}^{2}\sim r_{0}^2$.

The thermal-$AdS$ bulk action $S_{\rm{tbulk}}$ (\ref{oetbulkactionnn}) (after using the thermal-$AdS$ metric for $B\gg \Lambda_{IR}^{2}\sim r_{0}^2$ (\ref{thermalll}), evaluating the integrals, simplifying, and using $\beta'=\beta\sqrt{f}=\beta(1-\frac{1}{2}\frac{r_{h}^2}{r_{\Lambda}^2})$ near the boundary) becomes %and taking the $r_{\Lambda}\rightarrow \infty$ %limit gives the finite result
\begin{eqnarray}
S_{\rm{tbulk}}&=&3\beta\sqrt{f}B(r_{\Lambda}^2-r_{0}^2)\nonumber
\\
&=& 3\beta B(r_{\Lambda}^2-\frac{1}{2}r_{h}^2-r_{0}^2).
\end{eqnarray}
Hence,
\begin{eqnarray}\label{finalttt}
S_{\rm{thermal}}&=& S_{\rm{tbulk}} = 3\beta B(r_{\Lambda}^2-\frac{1}{2}r_{h}^2-r_{0}^2).
\end{eqnarray}

Therefore, $\Delta S_{\rm{E}}$ (which is the difference between the black hole (\ref{finalbhhh}) and thermal-$AdS_{5}$ (\ref{finalttt}) on-shell Euclidean actions) becomes
\begin{eqnarray}\label{change}
\Delta S_{\rm{E}}^{}&=& S_{\rm{bh}}-S_{\rm{thermal}}\nonumber
\\
&=& 3\beta B\left(r_{0}^2-\frac{r_{h}^2}{2}\right)\nonumber
\\
&=& 3\beta B\left(r_{0}^2-\frac{2}{9}\pi^2T^2\right),
\end{eqnarray}
for $B\gg T^{2}$.

For $B\gg T^{2}$, requiring $\Delta S_{\rm{E}}^{}(T_{c})=0$ (\ref{change}), we find the constraint equation for the critical temperature $T_{c}$ to be
\begin{equation}\label{}
 r_{0}^2-\frac{2}{9}\pi^2T_{c}^2=0,
\end{equation}
which can be solved for $T_{c}$ to give
\begin{equation}\label{largebtc}
  T_{c}(B, r_{0})=\frac{3}{\sqrt{2}}\frac{r_{0}}{\pi}.
\end{equation}

\bibliographystyle{prsty}
\bibliography{ar,tft,qft,books}

\end{document}